\documentclass[12pt]{article}
\usepackage{amssymb,amsmath,epsfig}
\allowdisplaybreaks
\begin{document}
\title{\bf Role of Curvature-Matter Coupling on Anisotropic Strange Stars}

\author{M. Sharif \thanks{msharif.math@pu.edu.pk} and Arfa Waseem
\thanks{arfawaseem.pu@gmail.com}\\
Department of Mathematics, University of the Punjab,\\
Quaid-e-Azam Campus, Lahore-54590, Pakistan.}
\date{}

\maketitle

\begin{abstract}
In this paper, we study the physical characteristics of anisotropic
spherically symmetric quark star candidates for $R+2\sigma T$
gravity model, where $R$, $\sigma$ and $T$ depict scalar curvature,
coupling parameter, and the trace of the energy-momentum tensor,
respectively. In order to analyze the structure formation of quark
stars, we consider the Heintzmann solution and assume that strange
quark matter is characterized by MIT bag model equation of state. We
evaluate the unknown parameters through matching conditions and
obtain the values of radii of strange quark stars using modified
Tolman-Oppenheimer-Volkoff equation with observed values of masses
and bag constant. The feasibility of our considered solution is
analyzed by graphical analysis of matter variables, energy bounds,
causality condition and adiabatic index. It is found that the
strange quark stars show stable structure corresponding to
Heintzmann solution and their physical viability enhances with
increasing values of the model parameter $\sigma$.
\end{abstract}

\section{Introduction}

The revolutionary developments in modern cosmology as well as
astrophysics provide a new inspiration about the universe and its
mysterious constituents. Despite the marvelous discoveries, there
are still various challenging problems which stimulate the
researchers to explore the hidden mysteries of these constituents
such as stars, galaxies and their respective physical phenomena.
Stars are mostly recognized as the basic ingredients of galaxy and
identified as the fundamental constituent in astronomy. The nuclear
fusion reactions in the core of a star exhibit a dynamical behavior
in the structure constitution as well as evolution of astrophysical
objects. The fusion processes in the star core generate outward
directed thermal pressure to overcome the gravitational pull
produced by a star mass. Once the nuclear fuel is entirely burnt
out, there remains no sufficient pressure to resist the attractive
force of gravity, consequently, the star experiences the stellar
death, a phenomenon known as gravitational collapse.

This collapse is responsible for the existence of massive stars
referred to as compact objects which are classified as white dwarfs,
neutron stars and black holes based on the masses of parent stars.
The study of interior configuration of compact stars has inspired
many observers to inspect their diverse features. In compact
objects, neutron stars have gained much attention due to their
compelling properties and structure compositions. In neutron stars,
the degeneracy pressure produced by the neutrons counterbalances the
gravitational pull to keep the star in a hydrostatic equilibrium
state. The structure of neutron stars with less dense core provides
an opportunity to transform into quark stars. The densest compact
objects composed of up, down and strange quark matter are named as
quark stars. This hypothetical compact object has drawn the
attention of many researchers to explore its interior formation
\cite{1}-\cite{10}.

In the analysis of structural properties of compact objects,
anisotropic matter configuration plays a crucial role. Since the
compact objects possess dense cores and have densities much larger
than the nuclear density, therefore, pressure should be anisotropic
in the interior region of celestial bodies \cite{11}. It is
anticipated that the interior constitution of quark stars is
expressed by MIT bag model equation of state (EoS) \cite{3,10}. Many
people examined the features of compact objects characterized by
this EoS along with anisotropic matter configuration. Demorest et
al. \cite{12} evaluated the mass of strange quark star (PSR
J1614-2230) and determined that such massive compact objects can
only be supported by MIT bag model EoS. Kalam et al. \cite{13}
considered a specific ansatz on the radial metric function to
examine the features of strange star candidates (SAX J 1808.4-3658,
Her X-1 and 4U 1820-30) for this EoS. They obtained non-singular
solutions of the field equations and found that their solutions
satisfy all the regularity conditions.

Rahaman et al. \cite{14} analyzed the presence of strange quark
stars using an interpolation function for mass and observed the
physical characteristics of stars associated with MIT bag model EoS.
For anisotropic distribution, Bhar \cite{15} established a new
hybrid star model by considering Krori-Barua ansatz with the same
EoS. She found that all the physical constraints are fulfilled and
the value of mass function is very close to the observational data.
Murad \cite{16} investigated the effects of charge on the
composition as well as nature of anisotropic quark stars.
Arba\~{n}il and Malheiro \cite{17} discussed the impact of
anisotropic factor on the stability of compact stars via numerical
solution of radial perturbation, the hydrostatic equilibrium and MIT
bag model. Deb and his collaborators \cite{18} obtained non-singular
anisotropic solutions for quark stars using the same EoS and
presented the graphical behavior of matter variables as well as
energy constraints for LMC X-4 star model.

General relativity (GR) has made remarkable achievements in
resolving various hidden puzzles of the cosmos but it is not
adequate to inspect the universe at large scales. In this regard,
alternative theories to GR are identified as the most effective
approaches to deal with the stimulating mysteries such as dark
matter and current accelerated cosmic expansion. This expansion
ensures the appearance of an enigmatic force with large negative
pressure dubbed as dark energy. Modified or extended theories of
gravity have shown a fundamental role to explore the diverse
features of dark energy and dark matter.

The simplest extension to GR is $f(R)$ theory \cite{19} originated
by placing a generic function $f(R)$ instead of curvature scalar
($R$) in the Einstein-Hilbert action. Other modified theories such
as $f(\mathcal{G})$ gravity ($\mathcal{G}$ shows Gauss-Bonnet
scalar) \cite{20} and $f(\mathcal{T})$ gravity ($\mathcal{T}$ stands
for torsion scalar) \cite{21} have achieved much attention due to
joined inspiration arising from cosmology, high-energy physics and
astrophysics. The fascinating feature of extended theories is based
on the coupling between matter and gravitational fields. Such type
of coupling provides an extra force which may yield interesting
results and helps to inspect hidden puzzles causing the cosmic
expansion. This leads to various modified theories having strong
curvature-matter coupling like $f(R,T)$ theory \cite{22}, $f(R,T,
R_{\gamma\delta}T^{\gamma\delta})$ theory \cite{23} and
$f(\mathcal{G},T)$ gravity \cite{24}.

The $f(R,T)$ theory as an extended form of $f(R)$ gravity has
inspired many researchers and is employed to narrate distinct
cosmological \cite{25}-\cite{34} and astrophysical
\cite{35}-\cite{43} scenarios. This theory has provided many
fascinating results in the evolution of compact objects associated
with MIT bag model EoS as well as anisotropic matter configuration.
In this context, Sharif and Siddiqa \cite{44} examined the influence
of curvature-matter coupling on the structure formation of compact
stars by using the MIT bag model and polytropic EoS for some
specific choices of the coupling parameters. Deb et al. \cite{45}
derived the analytic solutions for anisotropic strange stars with
same EoS and discussed the physical behavior of LMC X-4 as an
example of quark star. The influence of electric field on the
stability of Her X-1, 4U1820-30 and SAX J 1808.4- 3658 star
candidates are also investigated in this gravity \cite{46}. Biswas
et al. \cite{47} analyzed the anisotropic quark star models using
Krori-Barua solution with bag constant and observed the graphical
behavior of various physical quantities for three particular stars.

Recently, Deb et al. \cite{48} examined the properties of 12 strange
quark star candidates by assuming MIT bag model EoS for $R+2\sigma
T$ gravity model. They obtained the values of central energy density
and pressure for considered star models by employing Lake solution
and presented the graphical analysis for LMC X-4 star candidate. For
the same model, Maurya et al. \cite{48a} analyzed the behavior of
strange stars by adopting the embedding class one technique. In this
paper, we study the influence of anisotropic factor and MIT bag
model EoS on different strange star models by considering the
Heintzmann solution for $R+2\sigma T$ gravity model. The paper is
organized as follows. Next section deals with the formulation of
anisotropic $f(R,T)$ field equations corresponding to Heintzmann
ansatz. In section \textbf{3}, we derive the values of Heintzmann
constants by using boundary conditions at the surface of star.
Section \textbf{4} exhibits the graphical interpretation of physical
analysis of considered strange star models. Final remarks are given
in the last section.

\section{The $f(R,T)$ Field Equations}

The action of $f(R,T)$ theory associated with matter Lagrangian
($\mathcal{L}_{m}$) is specified by \cite{22}
\begin{equation}\label{1}
\mathcal{I}_{f(R,T)}=\int \left[\frac{f(R,
T)}{2\kappa}+\mathcal{L}_{m}\right]\sqrt{-g}d^{4}x,
\end{equation}
in which $\kappa=1$ stands for coupling constant and $g$ reveals the
determinant of the metric tensor ($g_{\gamma\delta}$). The
corresponding $f(R,T)$ field equations are
\begin{eqnarray}\nonumber
f_{R}(R,T)R_{\gamma\delta}&-&(\nabla_{\gamma}\nabla_{\delta}
-g_{\gamma\delta}\Box)f_{R}(R,T)-\frac{1}{2}g_{\gamma\delta}f(R,T)
\\\label{2}&=&T_{\gamma\delta}-(\Theta_{\gamma\delta}
+T_{\gamma\delta})f_{T}(R,T),
\end{eqnarray}
where $f_{R}(R,T)=\frac{\partial f(R,T)}{\partial
R},~f_{T}(R,T)=\frac{\partial f(R,T)}{\partial T}$,
$\Box=g^{\gamma\delta}\nabla_{\gamma}\nabla_{\delta}$,
$\nabla_{\gamma}$ is the covariant derivative and
$\Theta_{\gamma\delta}$ is characterized by
\begin{equation}\label{3}
\Theta_{\gamma\delta}=g^{\mu\upsilon}\frac{\delta
T_{\mu\upsilon}}{\delta g^{\gamma\delta}}=
g_{\gamma\delta}\mathcal{L}_{m}-2T_{\gamma\delta}-
2g^{\mu\upsilon}\frac{\partial^{2}\mathcal{L}_{m}}{\partial
g^{\gamma\delta}\partial g^{\mu\upsilon}}.
\end{equation}
The covariant divergence of Eq.(\ref{2}) provides
\begin{equation}\label{4}
\nabla^{\gamma}T_{\gamma\delta}=\frac{f_{T}}{1-f_{T}}
\left[(T_{\gamma\delta}+\Theta_{\gamma\delta})\nabla^{\gamma}(\ln
f_{T})-\frac{g_{\gamma\delta}}{2}\nabla^{\gamma}T+\nabla^{\gamma}
\Theta_{\gamma\delta}\right].
\end{equation}
This equation demonstrates that the energy-momentum tensor (EMT)
does not satisfy the conservation equation similar to other modified
theories \cite{22}.

In astrophysics, the matter configuration is characterized by EMT in
which all non-vanishing components express dynamical quantities
associated with some physical impacts. In the analysis of strange
compact structures, the anisotropy produced by pressure is
considered as a crucial ingredient which influences their structure
development. As the stellar distribution is mostly found to be
rotating as well as anisotropic therefore, the anisotropic parameter
has persuasive effects in different dynamical phases of stellar
transformation. The effect of anisotropy appears from the difference
of radial and transverse pressure components. Here, we discuss the
physical features of strange quark stars under the influence of
pressure anisotropy. We consider that the internal matter
distribution of stellar models is comprised by anisotropic matter
configuration whose mathematical description is
\begin{equation}\label{5}
T_{\gamma\delta}=(\rho+p_{t})U_{\gamma}U_{\delta}
-p_{t}g_{\gamma\delta}+(p_{r}-p_{t})X_{\gamma}X_{\delta},
\end{equation}
where $\rho$ indicates matter energy density, $p_{t}$ and $p_{r}$
are transverse and radial pressure entities, respectively,
$U_{\gamma}$ acts as four velocity and $X_{\gamma}$ denotes the
four-vector. The four velocity and four-vector in comoving
coordinates satisfy
\begin{equation}\nonumber
U^{\gamma}U_{\gamma}=1,\quad X_{\gamma}X^{\gamma}=-1.
\end{equation}
In the distribution of matter, there are various choices of
$\mathcal{L}_{m}$. Here, we assume $\mathcal{L}_{m}=-\mathcal{P}$,
where $\mathcal{P}=\frac{p_{r}+2p_{t}}{3}$ leading to
$\frac{\partial^{2}\mathcal{L}_{m}}{\partial
g^{\gamma\delta}\partial g^{\mu\upsilon}}=0$ \cite{22} and hence,
$\Theta_{\gamma\delta}=-2T_{\gamma\delta}-\mathcal{P}
g_{\gamma\delta}$. This form of matter Lagrangian has been used in
literature \cite{47}-\cite{48a} to analyze the nature of compact
objects.

To analyze the coupling outcomes of curvature and matter ingredients
in $f(R,T)$ scenario on the strange quark candidates, we choose an
independent model given as
\begin{equation}\label{6}
f(R,T)=f_{1}(R)+f_{2}(T).
\end{equation}
The $f(R,T)$ gravity includes $T$ that narrates more modified forms
of GR in comparison with $f(R)$ gravity. It is obvious that a
compatible as well as feasible model reveals the choices for
coupling constants whose values lie in the observational limits. We
consider $f_{1}(R)=R$ and $f_{2}(T)=2\sigma T$, where
$T=\rho-p_{r}-2p_{t}$. This particular model is firstly suggested by
Harko et al. \cite{22} and has been used extensively to study the
features of astrophysical objects \cite{49,50}. Recently, Deb et al.
\cite{48} found analytic solution of the generalized form of
Tolman-Oppenheimer-Volkoff (TOV) equation for this model and
discussed the features of various physical parameters for different
values of $\sigma$. Sharif and Siddiqa \cite{51} examined the
expanding as well as collapsing solutions corresponding to charged
cylindrical system and observed the viability of their solutions in
the same functional form.

On inserting the specific model along with the considered choice of
$\mathcal{L}_{m}$ in Eq.(\ref{2}), we obtain
\begin{equation}\label{7}
G_{\gamma\delta}=T_{\gamma\delta}+\sigma Tg_{\gamma\delta}
+2\sigma(T_{\gamma\delta}+\mathcal{P}g_{\gamma\delta})=T_{\gamma\delta}^{eff},
\end{equation}
where $G_{\gamma\delta}$ shows the standard Einstein tensor and
$T_{\gamma\delta}^{eff}$ represents effective EMT. For $\sigma=0$,
the standard results of GR can be retrieved from Eq.(\ref{7}). The
covariant divergence (\ref{4}) corresponding to $R+2\sigma T$ model
becomes
\begin{equation}\label{8}
\nabla^{\gamma}T_{\gamma\delta}=\frac{-\sigma}{1+2\sigma}
\big[g_{\gamma\delta}\nabla^{\gamma}T+2\nabla^{\gamma}
(\mathcal{P}g_{\gamma\delta})\big].
\end{equation}
In order to characterize the internal geometry of quark stars, we
take the static spherical stellar distribution expressed by the
metric
\begin{equation}\label{9}
ds^{2}_{-}=e^{\chi(r)}dt^{2}-e^{\eta(r)}dr^{2}
-r^{2}d\theta^{2}-r^{2}\sin^{2}\theta d\phi^{2}.
\end{equation}
Equations (\ref{5}), (\ref{7}) and (\ref{9}) lead to the following
field equations
\begin{eqnarray}\label{10}
\frac{1}{r^{2}}-e^{-\eta}\left(\frac{1}{r^{2}}-\frac{\eta'}{r}\right)&=&\rho
+\frac{\sigma}{3}(9\rho-p_{r}-2p_{t})=\rho^{eff},
\\\nonumber
e^{-\eta}\left(\frac{1}{r^{2}}+\frac{\chi'}{r}\right)-\frac{1}{r^{2}}&=&p_{r}
-\frac{\sigma}{3}(3\rho-7p_{r}-2p_{t})=p_{r}^{eff},
\\\label{11}\\\label{12}
e^{-\eta}\left(\frac{\chi''}{2}-\frac{\eta'}{2r}+\frac{\chi'}{2r}
+\frac{\chi'^{2}}{4}-\frac{\chi'\eta'}{4}\right)&=&p_{t}-\frac{\sigma}{3}
(3\rho-p_{r}-8p_{t})=p_{t}^{eff}.
\end{eqnarray}
Here prime reveals differentiation corresponding to $r$ and matter
variables $\rho^{eff}$, $p_{r}^{eff}$ and $p_{t}^{eff}$ indicate the
effective energy density, effective radial pressure and effective
transverse pressure of the compact system, respectively.

In order to solve the system (\ref{10})-(\ref{12}), we suppose that
inside the stellar models, physical quantities of the fluid
distribution are related through the MIT bag model EoS. This EoS
plays a dynamical role for the relativistic modeling of quark stars
\cite{3,10}. For massless as well as non-interacting strange quarks
matter in the bag model, the radial pressure is
\begin{equation}\label{13}
p_{r}=\sum_{\substack{q=u,d,s}}p^{q}-\mathfrak{B},
\end{equation}
with $p^{q}$ stands for the individual pressure corresponding to
each quark matter (up $(u)$, down $(d)$ and strange $(s)$) and
$\mathfrak{B}$ denotes the bag constant. The matter density with
respect to individual quark flavor is associated with the pressure
as $\rho^{q}=3p^{q}$. Consequently, the energy density for free
quarks in the bag model becomes
\begin{equation}\label{14}
\rho=\sum_{\substack{q=u,d,s}}\rho^{q}+\mathfrak{B}.
\end{equation}
Using Eqs.(\ref{13}) and (\ref{14}), the EoS for MIT bag model
describing the strange matter of quarks is derived as follows
\begin{equation}\label{15}
p_{r}=\frac{1}{3}(\rho-4\mathfrak{B}).
\end{equation}
To inspect the physical characteristics of quark star candidates,
many researchers have successfully used the MIT bag model EoS with
different choices of bag constant \cite{13}-\cite{18}. The study of
static spherically stellar system requires the following definition
of mass function
\begin{equation}\label{16}
m(r)=4\pi\int_{0}^{r}r^{2}\rho^{eff}dr.
\end{equation}
Insertion of Eq.(\ref{16}) into (\ref{10}) leads to
\begin{equation}\label{17}
e^{-\eta(r)}=1-\frac{2m(r)}{r},
\end{equation}
where $m$ represents the interior mass of a spherical object.

\subsection{Heintzmann Solution}

To examine the features of strange quark stars, we consider a
particular form of metric potentials proposed by Heintzmann
\cite{52} and defined as
\begin{equation}\label{18}
e^{\chi(r)}=\mathcal{A}^{2}(1+ar^{2})^{3}, \quad
e^{\eta(r)}=\left[1-\frac{3ar^{2}}{2}\left\{\frac{1+c(1
+4ar^{2})^{\frac{-1}{2}}}{1+ar^{2}}\right\}\right]^{-1},
\end{equation}
where $\mathcal{A}$, $a$ and $c$ are unknown constants. Heintzmann
provided this solution as the new Einstein static solution which may
generate interesting results in astrophysics. This solution is also
named as Heint IIa solution. Recently, Estrada and Tello-Ortiz
\cite{53} used this solution as an interior solution to obtain the
new anisotropic solution via gravitational decoupling approach. They
also analyzed the behavior of matter variables and stability
criterion for particular compact star models. Morales and
Tello-Ortiz \cite{54} also formulated charged anisotropic solutions
of the Einstein field equations by extending the Heintzmann solution
through gravitational decoupling technique.

Substituting Eq.(\ref{18}) in (\ref{10})-(\ref{12}) and using
(\ref{15}), we obtain
\begin{eqnarray}\label{19}
\rho^{eff}&=&\frac{-9a\left[(-3+ar^{2})(1+4ar^{2})^{\frac{3}{2}}
+c(-1+ar^{2}+14a^{2}r^{4})\right]}{4(1+2\sigma)(1+ar^{2})^{2}(1
+4ar^{2})^{\frac{3}{2}}}+\mathfrak{B},
\\\label{20}
p_{r}^{eff}&=&\frac{3a\left[(3-ar^{2})(1+4ar^{2})^{\frac{3}{2}}
+c(1-ar^{2}-14a^{2}r^{4})\right]}{4(1+2\sigma)(1+ar^{2})^{2}(1
+4ar^{2})^{\frac{3}{2}}}-\mathfrak{B},
\\\nonumber
p_{t}^{eff}&=&\frac{1}{2(1+2\sigma)(3+8\sigma)(1+ar^{2})^{2}(1
+4ar^{2})^{\frac{3}{2}}}\left[8\sigma(1+2\sigma)\sqrt{1+4ar^{2}}
\mathfrak{B}\right.\\\nonumber&+&\left.4a^{3}r^{4}\left\{\sqrt{1
+4ar^{2}}(2\sigma(-33+4\mathfrak{B}r^{2}(1+2\sigma))-27)-21c(3
+8\sigma)\right\}\right.\\\nonumber&+&\left.3a\left\{\sqrt{1+4a
r^{2}}(9+2\sigma(15+\mathfrak{B}r^{2}(1+2\sigma)))-c(3+2\sigma)\right\}
+3a^{2}r^{2}\right.\\\label{21}&\times&\left.\left\{\sqrt{1+4ar^{2}}(27
+2\sigma(49+12\mathfrak{B}r^{2}(1+2\sigma)))-c(33+70\sigma)\right\}\right].
\end{eqnarray}
For $\sigma=0$, these reduce to GR equations corresponding to MIT
bag model EoS. The mathematical form of anisotropic factor can be
evaluated as
\begin{eqnarray}\nonumber
\Delta^{eff}&=&p_{t}^{eff}-p_{r}^{eff}=\frac{3}{4(1+2\sigma)(3
+8\sigma)(1+ar^{2})^{2}(1+4ar^{2})^{\frac{3}{2}}}\Big[4(1+6\sigma
\\\nonumber&+&8\sigma^{2})\sqrt{1+4ar^{2}}
\mathfrak{B}+3a\Big\{\sqrt{1+4ar^{2}}(1+4\sigma)
(3+8\mathfrak{B}r^{2}(1+2\sigma))\\\nonumber&-&
c(3+4\sigma)\Big\}+3a^{2}r^{2}\Big\{\sqrt{1+4ar^{2}}(7
+36\sigma+12\mathfrak{B}r^{2}(1+6\sigma+8\sigma^{2}))
\\\nonumber&-&c(21+44\sigma)\Big\}+2a^{3}r^{4}\Big\{2
\sqrt{1+4ar^{2}}(-3(5+12\sigma)+4\mathfrak{B}r^{2}(1+6\sigma
\\\label{22}&+&8\sigma^{2}))\Big\}\Big].
\end{eqnarray}

\section{Boundary Conditions}

To examine the nature as well as physical features of anisotropic
strange quark stars, there must exist a smooth relationship between
both (interior and exterior) regions of compact objects. In this
regard, we choose the Schwarzschild metric presented by
\begin{equation}\label{23}
ds^{2}_{+}=\left(1-\frac{2\mathcal{M}}{r}\right)dt^{2}
-\left(1-\frac{2\mathcal{M}}{r}\right)^{-1}dr^{2}
-r^{2}d\theta^{2}-r^{2}\sin^{2}\theta d\phi^{2},
\end{equation}
where $\mathcal{M}$ acts as total mass within the boundary
$(r=\mathcal{R})$ of stellar model. At $r=\mathcal{R}$, the
continuity of metric coefficients $g_{tt}$, $g_{rr}$ and $g_{tt,r}$
leads to
\begin{eqnarray}\label{24}
1-\frac{2\mathcal{M}}{\mathcal{R}}&=&\mathcal{A}^{2}(1+a
\mathcal{R}^{2})^{3},\\\label{25}\left(1-\frac{2\mathcal{M}}
{\mathcal{R}}\right)^{-1}&=&\left[1-\frac{3a\mathcal{R}^{2}}{2}
\left\{\frac{1+c(1+4a\mathcal{R}^{2})^{-1/2}}{1+a\mathcal{R}^{2}}
\right\}\right]^{-1},\\\label{26}\frac{\mathcal{M}}{\mathcal{R}^{2}}
&=&3a\mathcal{A}^{2}\mathcal{R}(1+a\mathcal{R}^{2})^{2}.
\end{eqnarray}
The unknown triplet $(\mathcal{A},a,c)$ corresponding to radius and
total mass can be obtained from Eqs.(\ref{24})-(\ref{26}) which
provide
\begin{eqnarray}\label{27}
\mathcal{A}&=&\sqrt{\frac{(\mathcal{R}-2\mathcal{M})(3\mathcal{R}
-7\mathcal{M})^{3}}{\mathcal{R}(3\mathcal{R}-6\mathcal{M})^{3}}},
\\\label{28}a&=&\frac{\mathcal{M}}{\mathcal{R}^{2}(3\mathcal{R}
-7\mathcal{M})},\\\label{29}c&=&\left(\frac{3\mathcal{R}-8
\mathcal{M}}{\mathcal{R}}\right)\sqrt{\frac{3(\mathcal{R}
-\mathcal{M})}{3\mathcal{R}-7\mathcal{M}}}.
\end{eqnarray}

In $f(R,T)$ scenario, the modified form of generalized TOV equation
is
\begin{eqnarray}\label{30}
-p_{r}'-\frac{\chi'}{2}(\rho+p_{r})+\frac{2}{r}(p_{t}-p_{r})
+\frac{\sigma}{3(1+2\sigma)}(3\rho'-p_{r}'-2p_{t}')=0.
\end{eqnarray}
This equation along with Eqs.(\ref{11}) and (\ref{17}) provide the
hydrostatic equilibrium equation for anisotropic compact object in
$f(R,T)$ background as
\begin{eqnarray}\label{31}
p_{r}'=\frac{-\left[(\rho+p_{r})\left\{\frac{r^{2}}{2}p_{r}+\frac{m}{r}
-\frac{\sigma}{6}(3\rho-7p_{r}-2p_{t})\right\}\right]+2(p_{t}-p_{r})
\left(1-\frac{2m}{r}\right)}{r\left(1-\frac{2m}{r}\right)\left\{1
+\frac{\sigma}{3(1+2\sigma)}(1-\frac{3\rho'}{p_{r}'}+\frac{2p_{t}'}{p_{r}'})\right\}}.
\end{eqnarray}
This equation reduces to the standard hydrostatic equation in GR for
$\sigma=0$. Using condition $p_{r}(\mathcal{R})=0$ and
Eqs.(\ref{27})-(\ref{29}) in Eq.(\ref{20}), we obtain total mass of
the strange star as
\begin{equation}\label{31a}
\mathcal{M}=\frac{\mathcal{R}}{28}\Big[9+6\mathfrak{B}\mathcal{R}^{2}(1+2\sigma)
-\sqrt{\{9+6\mathfrak{B}\mathcal{R}^{2}(1+2\sigma)\}^{2}-336(1+2\sigma)
\mathfrak{B}\mathcal{R}^{2}}\Big].
\end{equation}
To evaluate the values of radii of the strange quark star models, we
solve Eq.(\ref{31}) using expressions given in (\ref{15}),
(\ref{19}) and (\ref{21}) for the observed values of masses of some
particular star candidates \cite{12}, \cite{54a}-\cite{54f} with
$\sigma=0.8$ and $\mathfrak{B}=64MeV/fm^{3}$. Table \textbf{1} gives
the values of predicted radii, compactness parameter and surface
gravitational redshift with respect to the masses of some strange
star candidates. It is interesting to mention here that the values
of these parameters are close to the values obtained by considering
different forms of metric functions in $f(R,T)$ framework for
$\mathfrak{B}=83MeV/fm^{3}$ \cite{48} and
$\mathfrak{B}=64MeV/fm^{3}$ \cite{48a}. Using the values of masses
and predicted radii of proposed strange stars, we evaluate the
values of Heintzmann constants ($\mathcal{A}, a, c$) given in Table
\textbf{2}.
\begin{table}
\caption{Physical values of the stellar models for
$\mathfrak{B}=64MeV/fm^{3}$ and $\sigma=0.8$.}
\begin{center}
\begin{tabular}{|c|c|c|c|c|}
\hline{Star Models}&{Mass $(M_{\odot})$}&{Predicted Radius
($km$)}&{$\frac{\mathcal{M}}{\mathcal{R}}$}&{$z_{s}$}\\
\hline{Her X-1}&{$0.85\pm0.15$}&{$7.64\pm0.20$}
&{0.163}&{0.218}\\
\hline{SAX J1808.4-3658}&{$0.9\pm0.3$}&{$8.05\pm0.71$}
&{0.164}&{0.219}\\
\hline{SMC X-1}&{$1.04\pm0.09$}&{$9.25\pm0.24$}&{0.165}
&{0.222}\\
\hline{LMC X-4}&{$1.29\pm0.05$}&{$10.58\pm0.11$}&{0.178}
&{0.246}\\
\hline{EXO 1785-248}&{$1.3\pm0.2$}&{$10.68\pm0.46$}&{0.179}
&{0.248}\\
\hline{Cen X-3}&{$1.49\pm0.08$}&{$11.06\pm0.31$}&{0.198}
&{0.287}\\
\hline{4U 1820-30}&{$1.58\pm0.06$}&{$11.70\pm0.12$}&{0.199}
&{0.289}\\
\hline{PSR J 1903+0327}&{$1.667\pm0.021$}&{$11.82\pm0.05$}
&{0.207}&{0.306}\\
\hline{Vela X-1}&{$1.77\pm0.08$}&{$12.08\pm0.10$}&{0.215}
&{0.324}\\
\hline{PSR J 1614-2230}&{$1.97\pm0.04$}&{$12.76\pm0.03$}
&{0.227}&{0.353}\\
\hline
\end{tabular}
\end{center}
\end{table}
\begin{table}
\caption{Calculated values of Heintzmann constants for different
star models.}
\begin{center}
\begin{tabular}{|c|c|c|c|}
\hline{Star Models}&{$\mathcal{A}$}&{$a$}&{$c$}\\
\hline{Her X-1}&{0.722669}&{$15.2\times10^{-4}$}&{1.9674}\\
\hline{SAX J1808.4-3658}&{0.721113}&{$14.3\times10^{-4}$}&{1.96198}\\
\hline{SMC X-1}&{0.719306}&{$10.5\times10^{-4}$}&{1.95569}\\
\hline{LMC X-4}&{0.691716}&{$9.2\times10^{-4}$}&{1.86018}\\
\hline{EXO 1785-248}&{0.692321}&{$9.0\times10^{-4}$}&{1.86226}\\
\hline{Cen X-3}&{0.653257}&{$10.0\times10^{-4}$}&{1.72859}\\
\hline{4U 1820-30}&{0.652265}&{$9.0\times10^{-4}$}&{1.72522}\\
\hline{PSR J 1903-327}&{0.633688}&{$9.6\times10^{-4}$}&{1.66224}\\
\hline{Vela X-1}&{0.616326}&{$9.9\times10^{-4}$}&{1.60367}\\
\hline{PSR J 1614-2230}&{0.590875}&{$9.8\times10^{-4}$}&{1.51825}\\
\hline
\end{tabular}
\end{center}
\end{table}

\section{Physical Features of Strange Quark Stars}

Here we analyze the physical properties of proposed anisotropic star
models. Using the calculated values of radii and Heintzmann
constants given in Tables \textbf{1} and \textbf{2}, respectively,
we study the behavior of various physical parameters. We examine the
viability of metric potentials and investigate the effective energy
density, effective radial and transverse pressure components,
anisotropic parameter, mass-radius relation, energy conditions and
stability for different values of the coupling parameter. It is
well-known that for a physically consistent solution, the metric
coefficients must be singularity free, monotonically increasing and
positive functions of $r$. The metric coefficients depend only on
Heintzmann constants as shown in Eq.(\ref{18}). Using the values of
these constants from Table \textbf{2} for some particular star
models, the behavior of metric functions is exhibited in Figure
\textbf{1} which shows that the metric functions are viable with all
the respective conditions and are physically consistent.
\begin{figure}\center
\epsfig{file=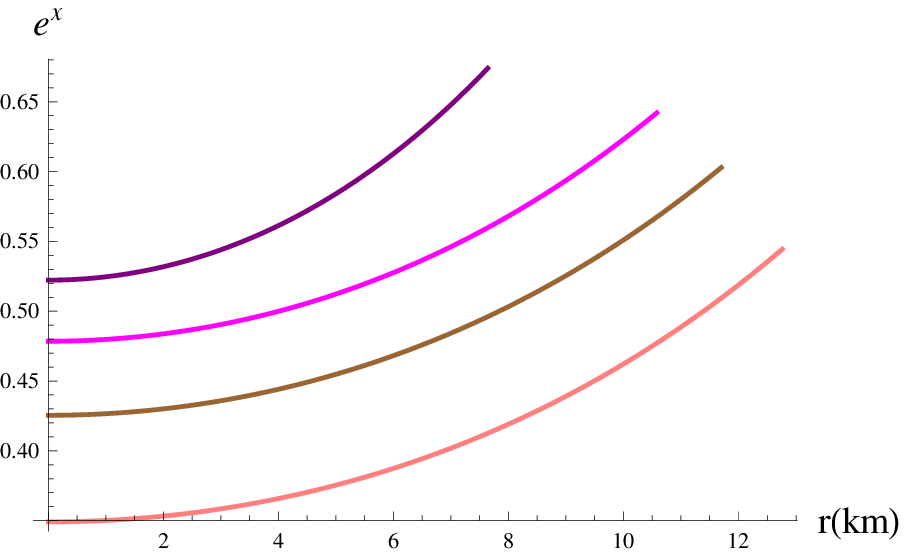,width=0.45\linewidth}
\epsfig{file=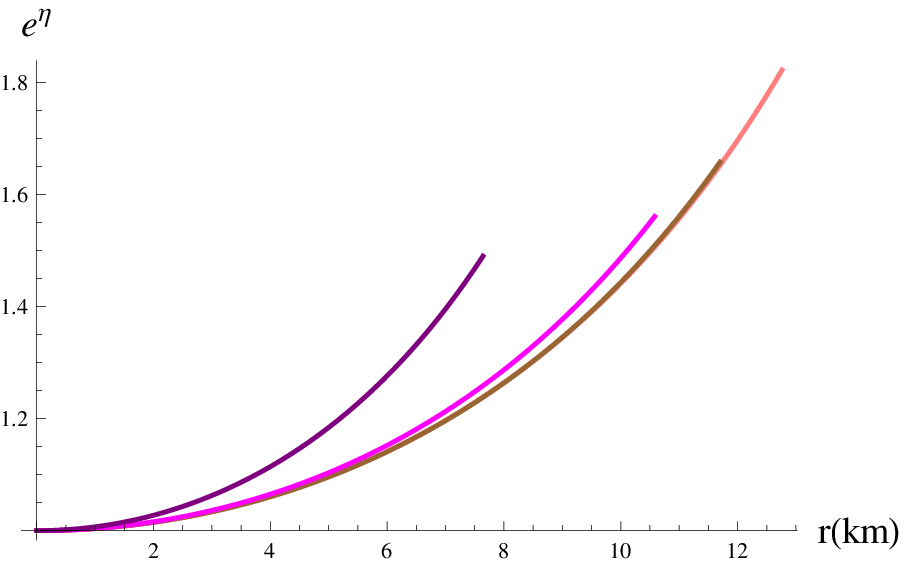,width=0.45\linewidth}\\
\caption{Behavior of metric potentials versus $r$ for Her X-1
(purple), LMC X-4 (magenta), 4U 1820-30 (brown) and PSR J 1614-2230
(pink).}
\end{figure}
\begin{figure}\center
\epsfig{file=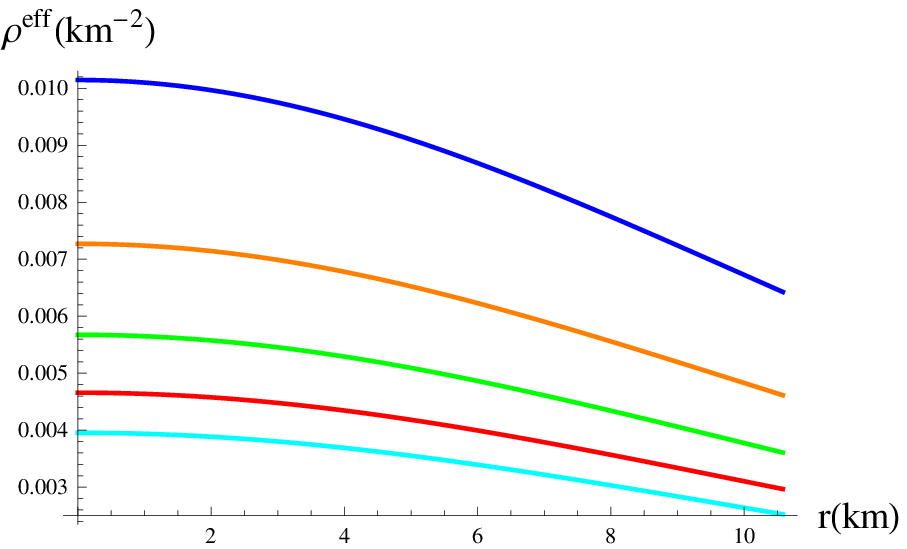,width=0.43\linewidth}
\epsfig{file=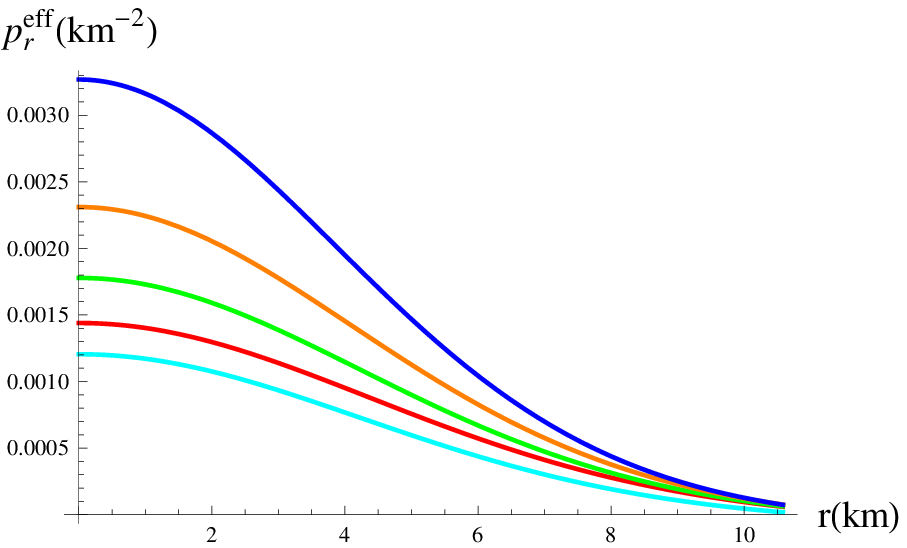,width=0.43\linewidth}
\epsfig{file=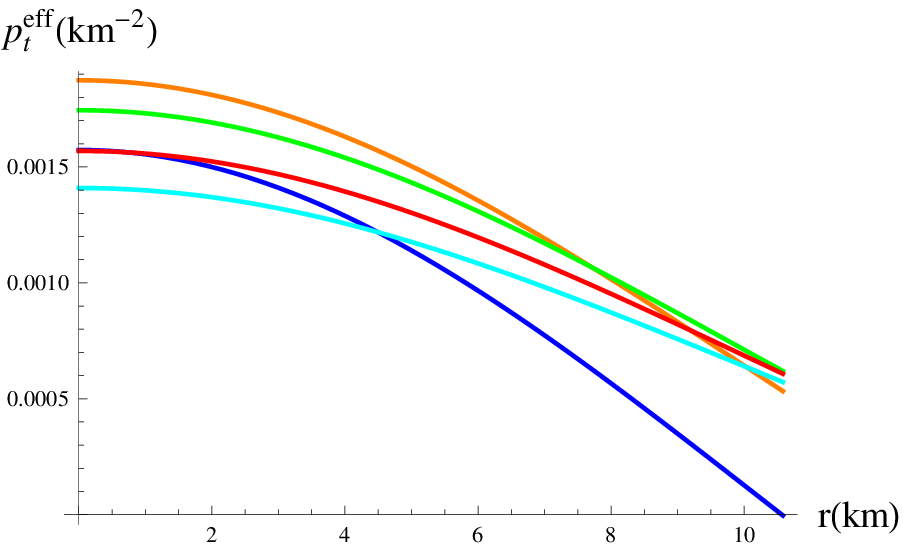,width=0.43\linewidth}
\epsfig{file=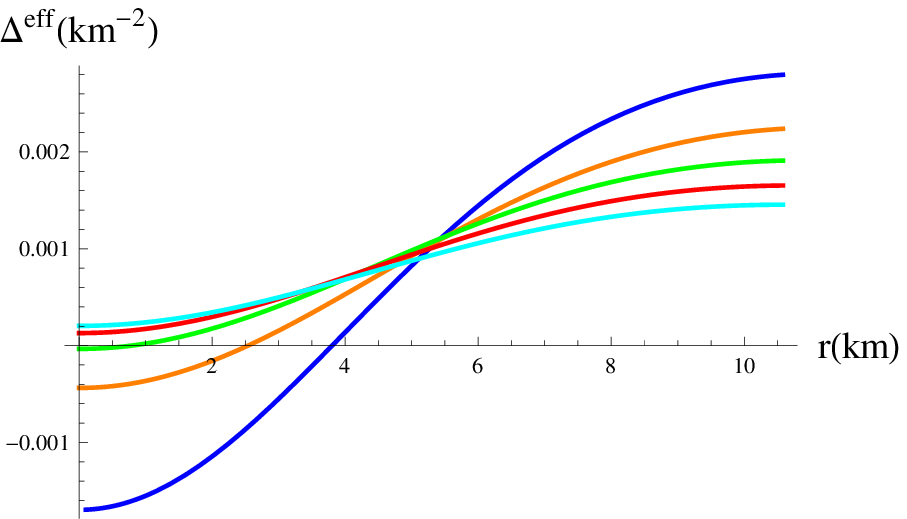,width=0.43\linewidth}\\
\caption{Variation of matter variables versus $r$ for LMC X-4 with
$\sigma=0$ (blue), $\sigma=0.2$ (orange), $\sigma=0.4$ (green),
$\sigma=0.6$ (red) and $\sigma=0.8$ (cyan).}
\end{figure}

\subsection{Analysis of Matter Components}

In the analysis of compact stellar objects, it is required that the
realistic compact structures should possess maximum pressure and
energy density in the core of a star. For different choices of
$\sigma$, the variation of effective energy density, effective
pressure (transverse/radial) and anisotropic component inside the
LMC X-4 star model is shown in Figure \textbf{2}. In these Figures,
$\sigma=0$ represents the standard GR results obtained in terms of
Heintzmann constants for $\mathfrak{B}=64MeV/fm^{3}$. We analyze the
graphical behavior of all the proposed star models but here we show
the graphical analysis only for LMC X-4 star model. Figure
\textbf{2} indicates that the effective energy density and effective
pressure components depict extreme values at the center of star.
These values show monotonically decreasing behavior towards the
boundary surface which confirm the presence of highly dense as well
as compact cores.
\begin{table}
\caption{Observed values of $\rho_{c}^{eff}$, $\rho_{0}^{eff}$ and
$p_{rc}^{eff}$ for $\sigma=0.8$ and $\mathfrak{B}=64MeV/fm^{3}$.}
\begin{center}
\begin{tabular}{|c|c|c|c|}
\hline{Star Models}&{$\rho_{c}^{eff} (gm/cm^{3})$}&{$\rho_{0}^{eff}
(gm/cm^{3})$}&{$p_{rc}^{eff} (dyne/cm^{2})$}\\
\hline{Her X-1}&{$8.98\times10^{15}$}&{$5.98\times10^{15}$}
&{$2.8\times10^{36}$}\\
\hline{SAX J1808.4-3658}&{$8.3\times10^{15}$}&{$5.4\times10^{15}$}
&{$2.6\times10^{36}$}\\
\hline{SMC X-1}&{$6.3\times10^{15}$}&{$4.1\times10^{15}$}
&{$1.9\times10^{36}$}\\
\hline{LMC X-4}&{$5.4\times10^{15}$}&{$3.5\times10^{15}$}
&{$1.6\times10^{36}$}\\
\hline{EXO 1785-248}&{$5.3\times10^{15}$}&{$3.4\times10^{15}$}
&{$1.6\times10^{36}$}\\
\hline{Cen X-3}&{$5.8\times10^{15}$}&{$3.6\times10^{15}$}
&{$1.6\times10^{36}$}\\
\hline{4U 1820-30}&{$5.1\times10^{15}$}&{$3.0\times10^{15}$}
&{$1.4\times10^{36}$}\\
\hline{PSR J 1903-327}&{$5.3\times10^{15}$}&{$3.4\times10^{15}$}
&{$1.5\times10^{36}$}\\
\hline{Vela X-1}&{$5.5\times10^{15}$}&{$3.1\times10^{15}$}
&{$1.6\times10^{36}$}\\
\hline{PSR J 1614-2230}&{$5.3\times10^{15}$}&{$2.9\times10^{15}$}
&{$1.5\times10^{36}$}\\
\hline
\end{tabular}
\end{center}
\end{table}

It is also observed that the value of radial pressure approximately
equals to zero at $r=\mathcal{R}$ for small values of $\sigma$ and
completely vanishes at $\sigma=0.8$. Moreover, the behavior of
anisotropic factor shows the variation from negative to positive
values with increasing values of $\sigma$. We can conclude that the
Heintzmann solution expresses the more viable behavior of
anisotropic factor and vanishing condition of radial pressure for
$\sigma=0.8$. Table \textbf{3} provides the values of effective
central energy density ($\rho_{c}^{eff}$), effective energy density
at the boundary surface ($\rho_{0}^{eff}$) and effective central
radial pressure ($p_{rc}^{eff}$) corresponding to $\sigma=0.8$ and
$\mathfrak{B}=64MeV/fm^{3}$ for all considered strange star models.
We note that these values of energy density and pressure are larger
than the values obtained for Lake solution in $f(R,T)$ scenario for
$\sigma=-0.4$ and $\mathfrak{B}=83MeV/fm^{3}$ \cite{48}.
\begin{figure}\center
\epsfig{file=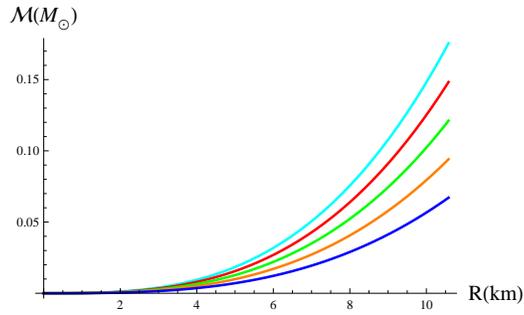,width=0.5\linewidth}\\
\caption{Variation of total mass versus total radius for LMC X-4
with $\sigma=0$ (blue), $\sigma=0.2$ (orange), $\sigma=0.4$ (green),
$\sigma=0.6$ (red) and $\sigma=0.8$ (cyan).}
\end{figure}

In Figure \textbf{3}, we present the behavior of total mass
normalized in units of solar mass with respect to the total radius
of the strange star candidates for different values of $\sigma$ and
the specific chosen value of bag constant. It is found that the mass
increases with increasing value of model parameter and the maximum
mass is $0.2M_{\odot}$ when $\sigma=0.8$. Hence, we conclude that
the increasing values of $\sigma$ lead to the more massive and less
dense compact objects.

\subsection{Energy Bounds}

The energy bounds have significant importance in illustrating the
existence of matter distribution. These bounds are utilized to
characterize the ordinary or exotic configuration of matter inside
the stellar objects as well as to examine the viability of the
theory. In $f(R,T)$ gravity for anisotropic fluid distribution,
these bounds are classified as \cite{55}
\begin{itemize}
\item Null : \quad$\rho^{eff}+ p_{r}^{eff}\geq 0$,
\quad$\rho^{eff}+ p_{t}^{eff}\geq 0$,
\item Weak:\quad$\rho^{eff}\geq 0$,\quad$\rho^{eff}
+ p_{r}^{eff}\geq 0$,\quad$\rho^{eff}+ p_{t}^{eff}\geq 0$,
\item Strong:\quad$\rho^{eff}+ p_{r}^{eff}\geq 0$,
\quad$\rho^{eff}+ p_{t}^{eff}\geq 0$,\quad$\rho^{eff} +
p_{r}^{eff}+2p_{t}^{eff}\geq 0$,
\item Dominant:\quad$\rho^{eff}-p_{r}^{eff}\geq0$,
\quad$\rho^{eff}- p_{t}^{eff}\geq0$.
\end{itemize}
Figure \textbf{4} shows that all energy bounds are fulfilled which
assure the presence of standard matter in strange star models. The
consistency of these bounds also indicates that our proposed
$f(R,T)$ model is physically acceptable for all chosen values of the
model parameter.
\begin{figure}\center
\epsfig{file=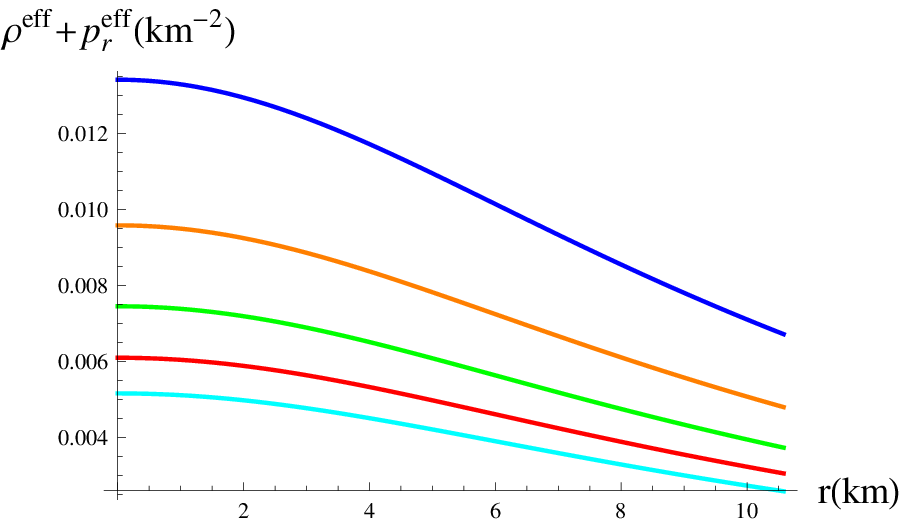,width=0.46\linewidth}
\epsfig{file=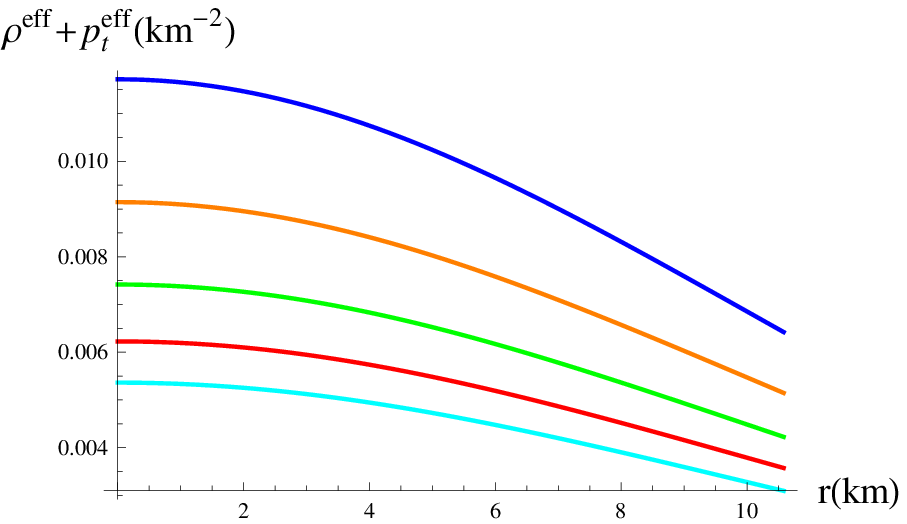,width=0.46\linewidth}
\epsfig{file=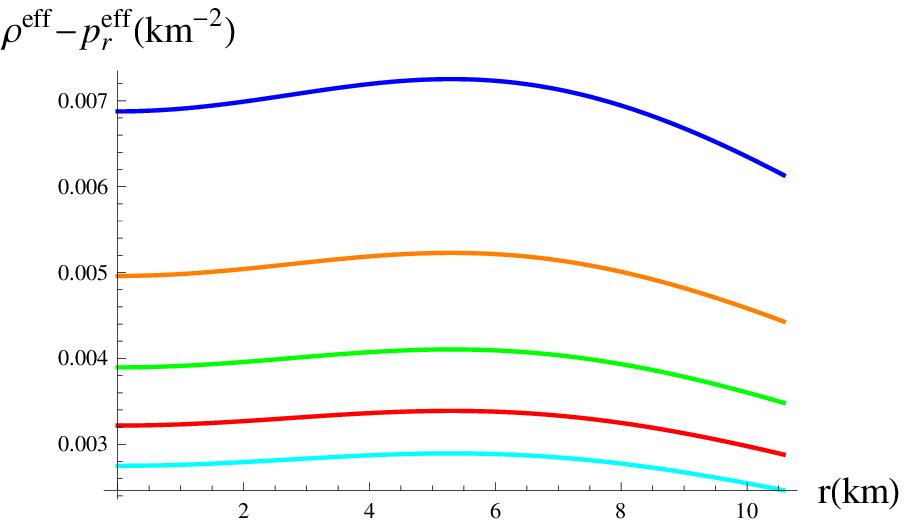,width=0.46\linewidth}
\epsfig{file=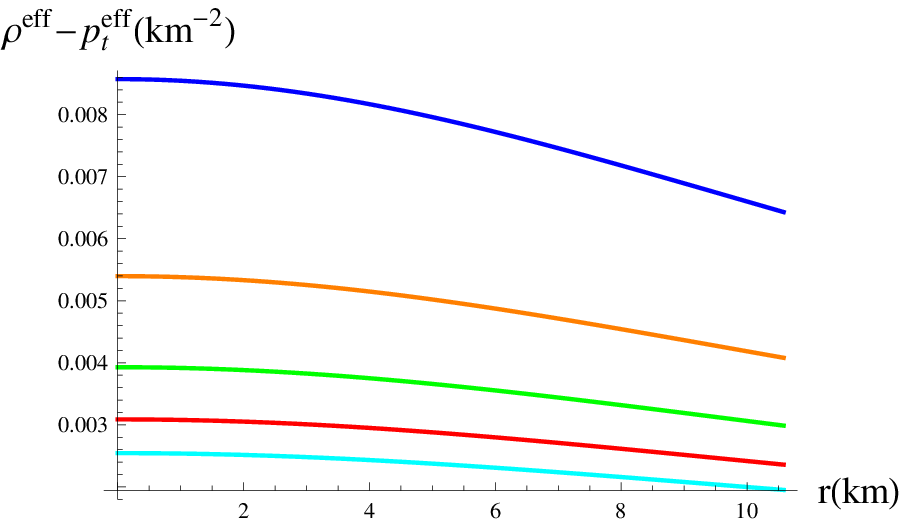,width=0.46\linewidth}
\epsfig{file=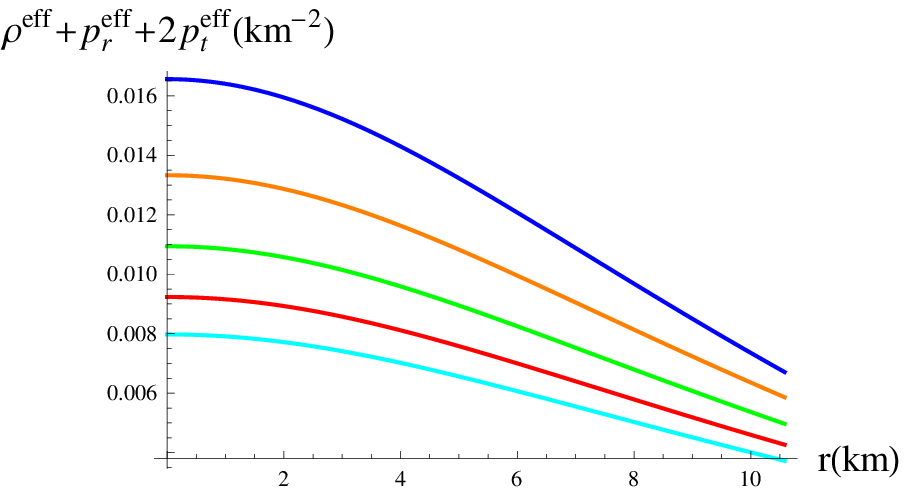,width=0.46\linewidth}\\
\caption{Energy bounds for LMC X-4 with $\sigma=0$ (blue),
$\sigma=0.2$ (orange), $\sigma=0.4$ (green), $\sigma=0.6$ (red) and
$\sigma=0.8$ (cyan).}
\end{figure}

\subsection{Compactness and Surface Redshift}

The ratio between mass and radius of a stellar object is known as
compactness factor. Using the expressions of Heintzmann constants
given in (\ref{27})-(\ref{29}), the mass of our stellar system is
evaluated as
\begin{equation}\label{32}
m(r)=\frac{3\mathcal{M}\left[1+\frac{3\mathcal{R}-8\mathcal{M}
\sqrt{\frac{3(\mathcal{R}-\mathcal{M})}{3\mathcal{R}-7\mathcal{M}}}}
{\mathcal{R}\sqrt{1+\frac{4\mathcal{M}r^{2}}{\mathcal{R}^{2}(3
\mathcal{R}-7\mathcal{M})}}}\right]r^{3}}{4\mathcal{R}^{2}\left[\mathcal{M}
\left(\frac{r^{2}}{\mathcal{R}^{2}}-7\right)+3\mathcal{R}\right]}.
\end{equation}
This clearly indicates that the mass function completely depends
upon the mass and radius of the strange stars. Equation (\ref{32})
also states that the mass function becomes zero at the center of
star, i.e., at $r=0$ and shows regular behavior within the interior
geometry of stellar structure. The compactness factor is defined by
\begin{equation}\nonumber
u=\frac{m(r)}{r}=\frac{3\mathcal{M}r^{2}\left[1+\frac{3\mathcal{R}
-8\mathcal{M}\sqrt{\frac{3(\mathcal{R}-\mathcal{M})}{3\mathcal{R}
-7\mathcal{M}}}}{\mathcal{R}\sqrt{1+\frac{4\mathcal{M}r^{2}}
{\mathcal{R}^{2}(3\mathcal{R}-7\mathcal{M})}}}\right]}{4
\mathcal{R}^{2}\left[\mathcal{M}\left(\frac{r^{2}}{\mathcal{R}^{2}}
-7\right)+3\mathcal{R}\right]}.
\end{equation}

The gravitational redshift ($z_{s}$) acts as a crucial parameter to
interpret the smooth relation between particles in the celestial
object and its EoS. In the framework of compactness parameter, the
gravitational redshift is expressed in the following form
\begin{equation}\nonumber
z_{s}=\frac{1}{\sqrt{1-\frac{3\mathcal{M}r^{2}\left[1+\frac{3
\mathcal{R}-8\mathcal{M}\sqrt{\frac{3(\mathcal{R}-\mathcal{M})}
{3\mathcal{R}-7\mathcal{M}}}}{\mathcal{R}\sqrt{1+\frac{4
\mathcal{M}r^{2}}{\mathcal{R}^{2}(3\mathcal{R}-7
\mathcal{M})}}}\right]}{2\left[\mathcal{M}
(r^{2}-7\mathcal{R}^{2})+3\mathcal{R}^{3}\right]}}}-1.
\end{equation}
Substituting the expression of $\mathcal{M}$ and the values of
radii, the behavior of compactness as well as redshift parameters is
obtained in Figure \textbf{5} for different values of $\sigma$.
These plots manifest that the required Buchdahl condition
$(\frac{\mathcal{M}}{\mathcal{R}}<\frac{4}{9})$ \cite{56} is
satisfied for compactness factor and the values of redshift
parameter are also in the desired range for anisotropic strange
star, i.e., $z_{s}\leq5.211$ \cite{57}.
\begin{figure}\center
\epsfig{file=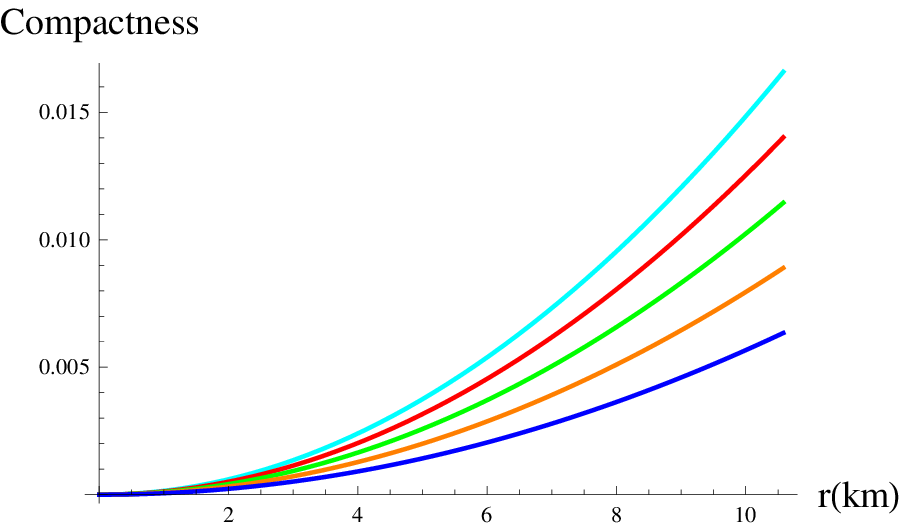,width=0.45\linewidth}
\epsfig{file=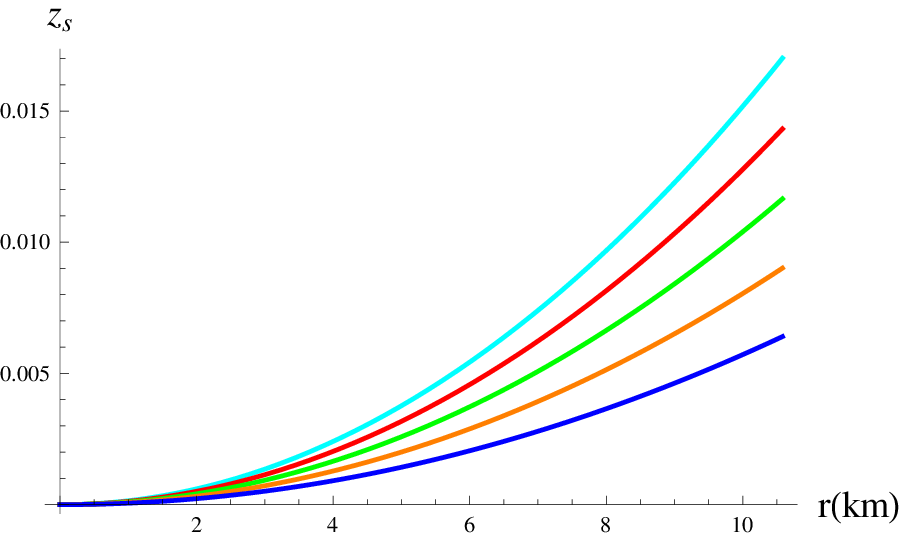,width=0.45\linewidth}\\
\caption{Variation of compactness parameter and surface redshift for
LMC X-4 with $\sigma=0$ (blue), $\sigma=0.2$ (orange), $\sigma=0.4$
(green), $\sigma=0.6$ (red) and $\sigma=0.8$ (cyan).}
\end{figure}

\subsection{Stability Analysis of Stellar Models}

The consistency of any physical system can be examined through the
stability of that system against external oscillations. The stellar
systems with stable structures are considered to be more realistic
in astrophysics. In order to investigate the stability of our
strange stars candidates, we check the causality condition and the
role of adiabatic index. In causality condition for anisotropic
fluid, the square of radial $(v^{2}_{sr})$ as well as transverse
($v^{2}_{st}$) speed of sounds should lie in the range [0, 1] for a
physically stable stellar object \cite{11}. Similarly, for a
potentially stable system based on Herrera's cracking concept, the
difference of radial and transverse sound speeds should possess the
same sign everywhere in the matter configuration, i.e., there should
be no cracking \cite{11}. Thus the causality as well as Herrera's
cracking concept demand that $0\leq v^{2}_{sr}\leq 1$, $0\leq
v^{2}_{st}\leq 1$ and $0\leq\mid v^{2}_{st}- v^{2}_{sr}\mid \leq 1$.
Here, the square sound speeds are expressed as
\begin{equation}\nonumber
v^{2(eff)}_{sr}=\frac{dp_{r}^{eff}}{d\rho^{eff}}, \quad
v^{2(eff)}_{st}=\frac{dp_{t}^{eff}}{d\rho^{eff}}.
\end{equation}
Figure \textbf{6} indicates that LMC X-4 star model has stable
structure with both causality and Herrera's cracking concept for the
specific values of $\sigma$ and bag constant.
\begin{figure}\center
\epsfig{file=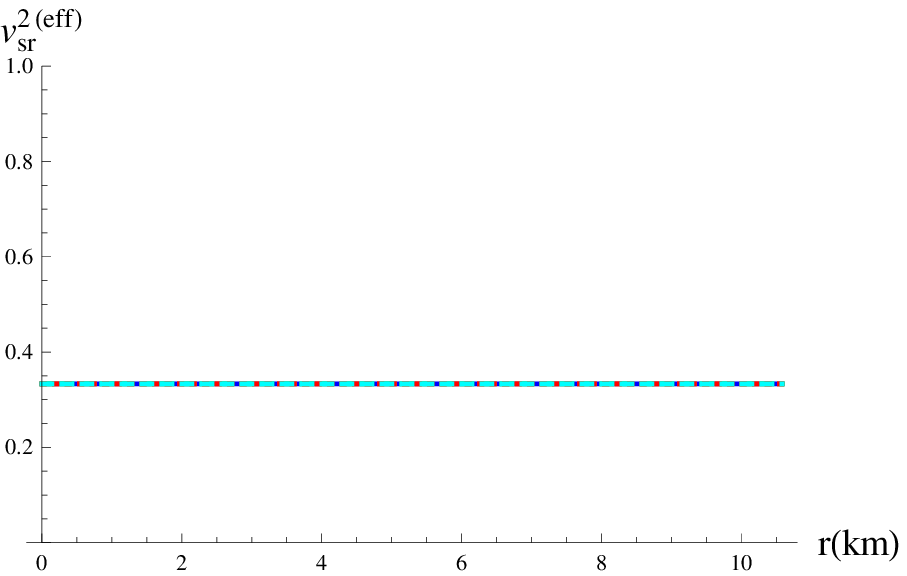,width=0.32\linewidth}
\epsfig{file=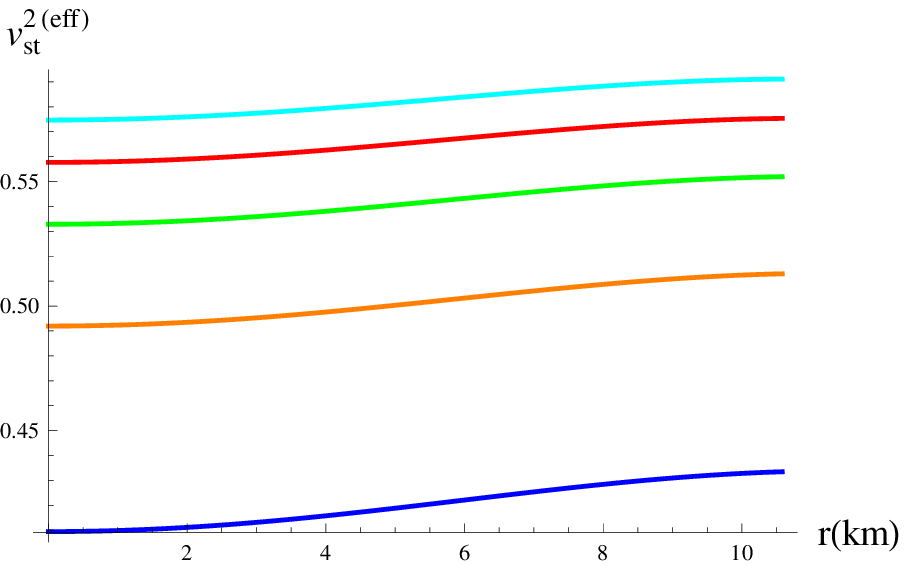,width=0.32\linewidth}
\epsfig{file=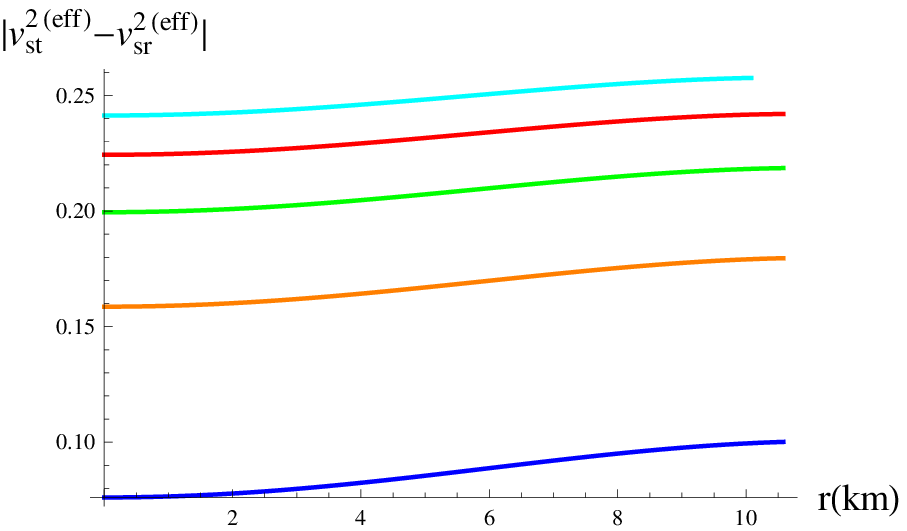,width=0.32\linewidth}\\
\caption{Analysis of stable structure for LMC X-4 with $\sigma=0$
(blue), $\sigma=0.2$ (orange), $\sigma=0.4$ (green), $\sigma=0.6$
(red) and $\sigma=0.8$ (cyan).}
\end{figure}

The validity of any EoS associated with the respective energy
density is given by the adiabatic index ($\Gamma$). This has gained
much attention to explain the stability of both relativistic and
non-relativistic stellar bodies. Following the work of Chandrasekhar
\cite{58}, researchers \cite{59,60} analyzed the dynamical stability
of compact objects against infinitesimal radial adiabatic
perturbation. It is proposed that for any dynamically stable
celestial object, the adiabatic index should be larger than
$\frac{4}{3}$ \cite{59}. For anisotropic matter distribution, the
mathematical form of adiabatic index is described as
\begin{equation}\nonumber
\Gamma=\frac{\rho^{eff}+p_{r}^{eff}}{p_{r}^{eff}}
\left(\frac{dp_{r}^{eff}}{d\rho^{eff}}\right).
\end{equation}
The graphical description of $\Gamma$ is presented in Figure
\textbf{7} for different values of $\sigma$. This demonstrates that
our proposed strange stars exhibit dynamically stable structure for
the considered choices of model parameter $\sigma$ and
$\mathfrak{B}$ as the value of $\Gamma$ $>\frac{4}{3}$ throughout
the system. Thus, the anisotropic stellar candidates show consistent
behavior corresponding to Heintzmann solution in $f(R,T)$ gravity.
\begin{figure}\center
\epsfig{file=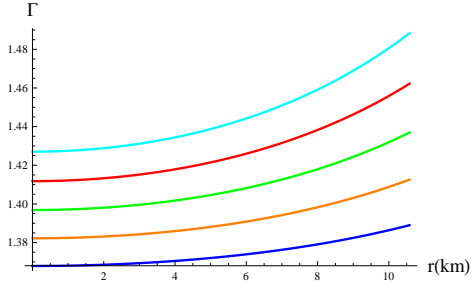,width=0.45\linewidth}\\
\caption{Role of adiabatic index versus $r$ for LMC X-4 with
$\sigma=0$ (blue), $\sigma=0.2$ (orange), $\sigma=0.4$ (green),
$\sigma=0.6$ (red) and $\sigma=0.8$ (cyan).}
\end{figure}

\section{Final Remarks}

This paper is devoted to discuss the influence of MIT bag constant
on the physical features of 10 anisotropic strange star candidates
in $f(R,T)$ gravity. We have employed the constraints on the metric
functions proposed by Heintzmann in which the arbitrary constants
($\mathcal{A}$, $a$, $c$) are determined by a linear relation
between both metrics of stellar objects. Using the observed values
of masses of considered stars in modified TOV equation, we have
evaluated the values of radii, compactness parameter and
gravitational redshift for $\mathfrak{B}=64 MeV/fm^{3}$ (Table
\textbf{1}) which are very close to the values obtained in
literature corresponding to the different choices of bag constant
\cite{48, 48a}.

We have obtained the central values of effective matter variables
(energy density and radial pressure) for proposed strange star
models through graphical analysis and showed their values for
$\sigma=0.8$ (Table \textbf{3}). We have presented the diagrammatic
interpretation of only LMC X-4 star model for different values of
$\sigma$. It is found that all the physical quantities indicate
positive, regular and finite behavior inside the strange stars that
reduce towards the surface of stellar models. For $\mathfrak{B}=64
MeV/fm^{3}$, the vanishing condition of the radial pressure is
satisfied at $r=\mathcal{R}$ for $\sigma=0.8$ and the behavior of
anisotropic factor is also found to be varying from negative to
positive values with increasing values of $\sigma$.

From the graphical interpretation of mass-radius relation of strange
star, it is obtained that with the larger values of model parameter,
the stellar system turns into a more massive and less dense compact
object. It is found that all energy bounds are fulfilled for
considered stellar models which guarantee the existence of ordinary
matter as well as the consistency of our $f(R,T)$ model. The maximum
values of the compactness and surface redshift are according to the
observational bounds. It is revealed that the stability constraints
are satisfied as the inequalities $0\leq v^{2(eff)}_{sr}\leq 1$,
$0\leq v^{2(eff)}_{st}\leq 1$ and $0\leq\mid v^{2(eff)}_{st}-
v^{2(eff)}_{sr}\mid \leq 1$ hold for LMC X-4 star which signifies
the existence of potentially stable constitution of quark stars. We
have also checked the stability in terms of adiabatic index and
observed that $\Gamma>\frac{4}{3}$ which indicates the stability
against an infinitesimal radial adiabatic perturbation.

It is observed that the matter-curvature coupling in $f(R,T)$
gravity may provide the more suitable results for ultra-dense
compact objects than GR \cite{48}. In this work, we have noticed
that strange stars depict regular as well as stable behavior in the
framework of Heinztmann solution for MIT bag model EoS and satisfy
all the required physical constraints when $\mathfrak{B}=64
MeV/fm^{3}$. In graphical analysis, we have presented the behavior
of all physical quantities for $\sigma=0$ which describes the
results of GR under the influence of Heintzmann constants. We
conclude that the vanishing condition of radial pressure and
positive behavior of anisotropic factor are obtained for
$\sigma=0.8$ rather than $\sigma=0$. Hence, our considered Heint IIa
solution along with the specific value of bag constant manifests
more viable structure of strange stars in $f(R,T)$ gravity as
compared to GR.

\vspace{.25cm}

{\bf Acknowledgment}

\vspace{0.25cm}

One (AW) of us would like to thank the Higher Education Commission,
Islamabad, Pakistan for its financial support through the {\it
Indigenous Ph.D. 5000 Fellowship Program Phase-II, Batch-III.}


\begin{thebibliography}{37}

\bibitem{1} Bodmer, A.R.: Phys. Rev. D \textbf{4}(1971)160.

\bibitem{2} Baym, G. and Chin, S.A.: Phys. Lett. B
\textbf{62}(1976)241.

\bibitem{3} Witten, E.: Phys. Rev. D \textbf{30}(1984)272.

\bibitem{4} Alcock, C., Farhi, E. and Olinto, A.: Astrophys. J.
\textbf{310}(1986)261.

\bibitem{5} Haensel, P., Zdunik, J.L. and Schaeffer, R.: Astron.
Astrophys. \textbf{160}(1986)121.

\bibitem{6} Li, X.D., Dai, Z.G. and Wang, Z.R.: Astron. Astrophys.
\textbf{303}(1995)L1.

\bibitem{7} Drago, A., Tambini, U. and Hjorth-Jensen, M.: Phys.
Lett. B \textbf{380}(1996)13.

\bibitem{8} Bombaci, I.: Phys. Rev. C \textbf{55}(1997)1587.

\bibitem{9} Dey, M. et al.: Phys. Lett. B \textbf{438}(1998)123.

\bibitem{10} Cheng, K.S., Dai, Z.G. and Lu, T.: Int. J. Mod. Phys. D
\textbf{7}(1998)139.

\bibitem{11} Herrera, L.: Phys. Lett. A \textbf{165}(1992)206.

\bibitem{12} Demorest, P.B. et al.: Nature \textbf{467}(2010)1081.

\bibitem{13} Kalam, M. et al.: Int. J. Theor. Phys. \textbf{52}(2013)3319.

\bibitem{14} Rahaman, F. et al.: Eur. Phys. J. C \textbf{74}(2014)3126.

\bibitem{15} Bhar, P.: Astrophys. Space Sci. \textbf{357}(2015)46.

\bibitem{16} Murad, M.H.: Astrophys. Space Sci. \textbf{361}(2016)20.

\bibitem{17} Arba\~{n}il, J.D.V. and  Malheiro, M.: J. Cosmol.
Astropart. Phys. \textbf{11}(2016)012.

\bibitem{18} Deb, D. et al.: Ann. Phys. \textbf{387}(2017)239.

\bibitem{19} Capozziello, S.: Int. J. Mod. Phys. D
\textbf{483}(2002)11; Nojiri, S. and Odintsov, S.D.: Phys. Rev. D
\textbf{68}(2003)123512; Carroll, S.M. et al.: Phys. Rev. D
\textbf{70}(2004)043528; Nojiri, S. and Odintsov, S.D.: Phys. Rev. D
\textbf{74}(2006)086005; Bertolami, O. et al.: Phys. Rev. D
\textbf{75}(2007)104016.

\bibitem{20} Cognola, G. et al.: Phys. Rev. D \textbf{73}(2006)084007;
Li, B., Barrow, J.D. and Mota, D.F.: Phys. Rev. D
\textbf{76}(2007)044027; Bamba, K. et al.: Eur. Phys. J. C
\textbf{67}(2010)295.

\bibitem{21} Bengochea, G.R. and Ferraro, R.: Phys. Rev. D
\textbf{79}(2009)124019; Linder, E.V.: Phys. Rev. D
\textbf{81}(2010)127301.

\bibitem{22} Harko, T. et al.: Phys. Rev. D \textbf{84}(2011)024020.

\bibitem{23} Haghani, Z. et al.: Phys. Rev. D \textbf{88}(2013)044023;
Odintsov, S.D. and S\'{a}ez-G\'{o}mes, D.: Phys. Lett. B
\textbf{725}(2013)437.

\bibitem{24} Sharif, M. and Ikram, A.: Eur. Phys. J. C
\textbf{76}(2016)640.

\bibitem{25} Sharif, M. and Zubair, M.: J. Cosmol. Astropart. Phys.
\textbf{03}(2012)028.

\bibitem{26} Sharif, M. and Zubair, M.: J. Phys. Soc. Jpn.
\textbf{81}(2012)114005.

\bibitem{27} Jamil, M., Momeni, D. and Myrzakulov, R.: Chin.
Phys. Lett. \textbf{29}(2012)109801.

\bibitem{28} Sharif, M. and Zubair, M.: J. Phys. Soc. Jpn.
\textbf{82}(2013)064001; ibid. 014002.

\bibitem{29} Shabani, H. and Farhoudi, M.: Phys. Rev. D
\textbf{88}(2013)044048.

\bibitem{30} Sharif, M. and Zubair, M.: Gen. Relativ. Gravit.
\textbf{46}(2014)1723.

\bibitem{31} Shabani, H. and Farhoudi, M.: Phys. Rev. D
\textbf{90}(2014)044031.

\bibitem{32} Moraes, P.H.R.S.: Eur. Phys. J. C \textbf{75}(2015)168.

\bibitem{33} Momeni, D., Myrzakulov, R. and G\"{u}dekli, E.:
Int. J. Geom. Meth. Mod. Phys. \textbf{12}(2015)1550101.

\bibitem{34} Moraes, P.H.R.S.: Int. J. Theor. Phys. \textbf{55}(2016)1307.

\bibitem{35} Alhamzawi, A. and Alhamzawi, R.: Int. J. Mod. Phys. D
\textbf{25}(2015)1650020.

\bibitem{36} Moraes, P.H.R.S., Arba\~{n}il, J.D.V. and
Malheiro, M.: J. Cosmol. Astropart. Phys. \textbf{06}(2016)005.

\bibitem{37} Zubair, M., Abbas, G. and Noureen, I.:
Astrophys. Space Sci. \textbf{361}(2016)8.

\bibitem{38} Sharif, M. and Siddiqa, A.: Eur. Phys. J. Plus
\textbf{132}(2017)529.

\bibitem{39} Das, A.: Phys. Rev. D \textbf{95}(2017)124011.

\bibitem{40} Yousaf, Z., Bhatti, M.Z. and Ilyas, M.:
Eur. Phys. J. C \textbf{78}(2018)307.

\bibitem{41} Sharif, M. and Waseem, A.: Gen. Relativ. Gravit.
\textbf{50}(2018)78.

\bibitem{42} Deb, D. et al.: J. Cosmol. Astropart. Phys.
\textbf{03}(2018)044.

\bibitem{43} Sharif, M. and Waseem, A.: Eur. Phys. J. C
\textbf{50}(2018)78.

\bibitem{44} Sharif, M. and Siddiqa, A.: Int. J. Mod. Phys. D
\textbf{27}(2018)1850065.

\bibitem{45} Deb, D. et al.: Phys. Rev. D \textbf{97}(2018)084026.

\bibitem{46} Sharif, M. and Waseem, A.: Int. J. Mod. Phys. D
\textbf{28}(2019)1950033.

\bibitem{47} Biswas, S. et al.: Ann. Phys. \textbf{401}(2019)1.

\bibitem{48} Deb, D. et al.: Mon. Not. R. Astron. Soc. \textbf{485}(2019)5652.

\bibitem{48a} Maurya, S.K. et al.: Phys. Rev. D \textbf{100}(2019)044014.

\bibitem{49} Das, A. et al.: Eur. Phys. J. C \textbf{76}(2016)654.

\bibitem{50} Sharif, M. and Siddiqa, A.: Eur. Phys. J. Plus
\textbf{133}(2018)226.

\bibitem{51} Sharif, M. and Siddiqa, A.: Ad. High Energy Phys.
\textbf{2019}(2019)8702795.

\bibitem{52} Heintzmann, H.: Z. Phys. \textbf{228}(1969)489.

\bibitem{53} Estrada, M. and Tello-Ortiz, F.: Eur. Phys. J.
Plus \textbf{133}(2018)453.

\bibitem{54} Morales, E. and Tello-Ortiz, F.: Eur. Phys. J. C
\textbf{78}(2018)618.

\bibitem{54a} Abubekerov, M.K. et al.: Astron. Rep.
\textbf{52}(2008)379.

\bibitem{54b} Elebert, P. et al.: Mon. Not. R. Astron. Soc.
\textbf{395}(2009)884.

\bibitem{54c} \"{O}zel, F.,  G\"{u}ver, T. and Psaltis, D.:
Astrophys. J. \textbf{693}(2009)1775.

\bibitem{54d} G\"{u}ver, T. et al.: Astrophys. J. \textbf{719}(2010)1807.

\bibitem{54e} Rawls, M. et al.: Astrophys. J. \textbf{730}(2011)25.

\bibitem{54f} Freire, P.C.C. et al.: Mon. Not. R. Astron. Soc.
\textbf{412}(2011)2763.

\bibitem{55} Chakraborty, S.: Gen. Relativ. Gravit.
\textbf{45}(2013)2039.

\bibitem{56} Buchdahl, H.A.: Phys. Rev. \textbf{116}(1959)1027.

\bibitem{57} Ivanov, B.V.: Phys. Rev. D \textbf{65}(2002)104011.

\bibitem{58} Chandrasekhar, S.: Astrophys. J. \textbf{140}(1964)417.

\bibitem{59} Heintzmann, H. and Hillebrandt, W.: Astron. Astrophys.
\textbf{38}(1975)51.

\bibitem{60} Hillebrandt, W. and Steinmetz, K.O.: Astron. Astrophys.
\textbf{53}(1976)283; Bombaci, I.: Astron. Astrophys.
\textbf{305}(1996)871.


\end{thebibliography}
\end{document}